# A Study of Data Pre-processing Techniques for Imbalanced Biomedical Data Classification

## Shigang Liu*; Jun Zhang*; Yang Xiang; Wanlei Zhou; Dongxi Xiang

Address:
Department of Computer Science and Software Engineering, Swinburne University of Technology, Hawthorn, VIC 3122, Australia, email: shigangliu@swin.edu.au
Department of Computer Science and Software Engineering, Swinburne University of Technology, Hawthorn, VIC 3122, Australia, email: junzhang@swin.edu.au
Department of Computer Science and Software Engineering, Swinburne University of Technology, Hawthorn, VIC 3122, Australia, yxiang@swin.edu.au
School of Information Technology, Deakin University, Burwood, VIC 3125, Australia, email: wanlei@deakin.edu.au
Department of Genetics, Harvard Medical School, Boston MA 02115, American, email: dxiangmedsci@gmail.com

Biographical notes:
Shigang Liu is a research fellow with School of Software and Electrical Engineering, Swinburne University of Technology. His research interests include applied machine learning, data analytics and so on.
Jun Zhang is an Associate Professor with School of Software and Electrical Engineering, and the Deputy Director of Swinburne Cybersecurity Lab, Swinburne University of Technology, Australia. His research interests include cybersecurity and applied machine learning.
Yang Xiang is the Dean of Digital Research & Innovation Capability Platform, Swinburne University of Technology, Australia. His research interests include data analytics, cyber security and so on.
Wanlei Zhou is currently the Alfred Deakin Professor (the highest honour the University can bestow on a member of academic staff) and Associate Dean- International Research Engagement, Deakin University. His research interests include bioinformatics, E-learning and so on.
Dongxi Xiang is a Research Fellow with Department of Genetics, Harvard Medical School. His research interests include Breast Cancer, Cancer Stem Cell, Aptamer and Targeted Cancer Therapy.

*Corresponding author

**Abstract**— Biomedical data are widely accepted in developing prediction models for identifying a specific tumor, drug discovery and classification of human cancers. However, previous studies usually focused on different classifiers, and overlook the class imbalance problem in real-world biomedical datasets. There are a lack of studies on evaluation of data pre-processing techniques, such as resampling and feature selection, on imbalanced biomedical data learning. The relationship between data pre-processing techniques and the data distributions has never been analysed in previous studies. This article mainly focuses on reviewing and evaluating some popular and recently developed resampling and feature selection methods for class imbalance learning. We analyse the effectiveness of each technique from data distribution perspective. Extensive experiments have been done based on five classifiers, four performance measures, eight learning techniques across twenty real-world datasets. Experimental results show that: (1) resampling and feature selection techniques exhibit better performance using support vector machine (SVM) classifier. However, resampling and Feature Selection techniques perform poorly when using C4.5 decision tree and Linear discriminant analysis classifiers; (2) for datasets with different distributions, techniques such as Random undersampling and Feature Selection perform better than other data pre-processing methods with T Location-Scale distribution when using SVM and KNN (K-nearest neighbours) classifiers. Random oversampling outperforms other methods on Negative Binomial distribution using Random Forest classifier with lower level of imbalance ratio; (3) Feature Selection outperforms other data pre-processing methods in most cases, thus, Feature Selection with SVM classifier is the best choice for imbalanced biomedical data learning.

# 1 INTRODUCTION

Gene expression profiling has been one of the most important molecular biology technologies in post-genomic era (Yu et al. 2013). It is successfully used in the development of class prediction models for identifying a specific tumor (Wigle et al. 2002), prognostics and disease diagnostics (Golub et al. 1999), (Nutt et al. 2003), (Khoshgoftaar et al. 2014), (Conrads et al. 2003). However, a well-known fact is that real-world data in general have class imbalance problem, where the samples of one class outnumber the samples of other class(es) (Yang and Wu 2006).When facing unequal distribution of training data, traditional classifiers are often biased toward the majority class and perform poorly with the minority class. Traditional machine learning algorithms are desired to maximize overall number of correct predictions without giving enough consideration of the minority examples. For example, given a dataset, where five percent and ninety-five percent are minority class samples and majority class samples, respectively. If a classifier recognizes all data as the majority class, the classification accuracy would be ninety-five percent. However, this classifier is not useful in practice. For many real-world problems, the class of interest is the minority class. How to accurately identify the minority class sample is a more challenging problem. The problem has drawn significant interest since the year 2000 from data mining, knowledge discovery, machine learning and artificial intelligence. Imbalanced learning has become an especially hot topic under some special issues (Yu et al. 2013), conference and workshops such as BOSC'15 (Fogg and Kovats 2015), WABI'15 (Pop and Touzet 2015), BIBM'14 (Liang et al. 2015), GIW'14 (Shibuya et al. 2015). Class imbalance is also a common problem with biomedical data (Liu et al. 2009), (Yu and Ni 2014), (Lin and Chen 2013). The impact of class imbalance in biomedical data could be even worse. For instance, if a potential cancer patient was predicted as non-cancer, the patient could loss his/her life because of the delay in the correct diagnosis and treatment. Therefore, it is very necessary to deal with the class imbalance problem and significantly improve the accuracy of a classification model.

Class imbalance problem has been studied by many researchers (Yu et al. 2014), (Lin et al. 2013), (Rahman and Davis

2013), (Khoshgoftaar et al. 2015), (Lusa 2010). Methods for addressing class imbalance problem mainly include three categories: resampling (Rahman et al. 2013), (Lusa 2010), (He et al. 2008), (Barandela et al. 2004), (Jo and Japkowicz 2004), cost-sensitive learning (Provost and Fawcett 2001), (Sun et al. 2007), (Zhou and Liu 2006), (Liu and Zhou 2006), (Eitrich et al. 2007) and ensemble learning (Yu et al. 2014), (Lin et al. 2013), (Chawla et al. 2003), (Tao et al. 2006), (Li et al. 2008), (Khoshgoftaar et al. 2011). For example, Khoshgofta et al. (Khoshgoftaar et al. 2015) investigated the class imbalance problem which is mainly focused on random undersampling, Select-Bagging and Select-Boosting. Sun et al. have studied the classification of imbalanced data from cost-sentive boosting perspective with respect to their weighting strategies towards different types of samples (Sun et al. 2007); López et al. presented a comparative study about preprocessing and cost-sensitive learning when dealing with imbalance using two oversampling methods, a cost-sensitive version and a hybrid approach (López et al. 2012); Lin and Chen (Lin et al. 2013) presented a comparative study using five genomic datasets and four classifiers, with each coupled with an ensemble correction strategy and one support vector machines (SVM)-based classifier. Recently, feature selection method has achieved outstanding performance in addressing high-dimensional imbalanced biomedial data (Tiwari 2014), (Yin et al. 2013), (Yu et al. 2014). However, none of the previous studies have specifically studied the resampling methods on biomedical data. Therefore, we believe it is highly essential to further explore the resampling technique, not only because it is one of the most popular class imbalance learning techniques, but also due to the fact that recently developed techniques such as CBUS and feature selection [8] have not been investigated in previous studies. An effective comparison of resampling with feature selection methods across biomedical datasets have not been conducted before. Most importantly, data distribution has never been considered in previous imbalanced biomedical data studies.

Different from other related work, the experimentation study in this paper mainly focuses on resampling and feature selection techniques in class imbalance problem with data distribution being considered as well. The main contributions of this paper are as follows: (1) we have conducted an extensive experiment study and (2) the relationship between data distributions and different class imbalance learning techniques have been discussed. Precisely, for the former contribution: firstly, our study focuses on recently developed and popularly used sampling techniques. In the meantime, considering that feature selection (FS) is also beneficial to imbalanced data learning, one of the recently developed FS approaches is also employed in this study (Yu et al. 2014). Secondly, five classification algorithms have been considered in the study, therefore, the experimental results are more convictive, extensive and comparable. Thirdly, an analysis of each technique regarding different classification algorithms have been provided. Thus, communities in bioinformatics can choose a specific resampling approach or the feature selection method with respect to different classifiers for a given learning scenario. For the latter contribution, we have identified the data distribution of each dataset used in this study, and a further analysis of each class imbalance learning technique regarding different data distribution has been reported. The experimental results are very attractive, in this case communities in bioinformatics area can choose an effective technique once the distribution of the data is known in advance. To the best of our knowledge, no previous comprehensive empirical investigations have been performed in comparing the performance of imbalanced data learning methods with data distribution being considered.

The rest of the paper is organized as follows. Section II introduces the techniques to be evaluated in this study, while the details of datasets are presented in Section III. The experimental design, performance metrics and classification algorithms are outlined in Section IV. Section V discusses the experimental result. The conclusion is provided in Section VI.

## 2 Methods to Be Evaluated

Generally speaking, approaches to classification with imbalanced data issues involve three main categories: resampling, cost-sensitive methods and the ensemble methods (He and Garcia 2009), (Yang et al. 2014), (Lin et al. 2013). In this subsection we only review the techniques that to be evaluated in this study. For the detailed information of each technique, please refer the related work (He et al. 2009), (Yang et al. 2014), (Lin et al. 2013).

**Random over-sampling (ROS):** In ROS, new minority samples are created by randomly selecting training samples from minority class, and then duplicating it. In doing so, the class distribution can be balanced, but this may usually cause over-fitting and longer training time during imbalance learning process.

**Random under-sampling (RUS):** This technique draws a random subset from the majority class while discarding the rest of instances, where the class distribution can be balanced. The size of the subset is calculated according to the desired class distribution ratio. However, one common criticism is that some important information may be lost when examples are removed from the training dataset, especially for a small dataset.

**Wilson's Editing(WE) (Barandela et al. 2004):** WE which was introduced by Barandela et al., modifies the older strategy (by Wilson) for pruning a data set in order to improve the balance level. Precisely, this technique consists of applying the $k$NN classifier with $k = 3$ to classify each example in the training set by using all the remaining examples, and removing those majority class instances whose class label does not agree with the class associated with the largest number of the $k = 3$ neighbours. Realized that the editing technique did not produce significant reductions in the size of the majority class. Barandela et al. modified the distance calculation with the weighted distance below mentioned, which has taken the class into account. Formula (1) is the modified weighting distance, where $N_i$ is the number of examples in class $i$ of the training data, $N$ is the total number of samples in the dataset, and $m$ is the number of features in each sample. We can see that the weighting distance for a minority class sample is smaller than the weighting distance for a majority class sample.

$$\text{weighting distance} = \left(\frac{N_i}{N}\right)^{1/m} \tag{1}$$

**Cluster-based over-sampling(CBOS) (Jo et al. 2004):** The main idea of this approach is that before performing random over-sampling, first using k-means algorithm to cluster the minority and majority classes separately. After the training examples of each class have been clustered, all clusters in the majority class, except for the largest one, are randomly over-sampling as the same number of the training examples as the largest cluster samples. Then the total number of the majority clusters are even out to each cluster of minority clusters. Thus, the minority class and majority class are balanced with the same number of examples. Finally, merge the updated minority class and majority class into one data set as a new training data set. Take the example from (Jo et al. 2004), let assume that the training examples of the minority and majority classes are respectively clustered as follows:

Majority class: 10, 10, 10, 24 (which means there are four clusters with each cluster has 10, 10, 10, 24 examples, respectively).

Minority class: 2, 3, 2

According to CBOS, we obtain the below new distribution of each cluster:

Majority class: 24, 24, 24, 24

Minority class: 32, 32, 32

That is to way, in the majority class, all size 10 clusters are oversampled to 24 training samples, which is the largest majority subcluster. In this respect, the minority should have the same number of examples which is 96 after resampling, since it includes only three clusters, therefor, each minority class cluster is randomly oversampled until it contains $96/3 = 32$ examples.

**Cluster-based under-sampling (CBUS) (Rahman et al. 2013):** The aim of Cluster based under-sampling approach is not to balance the data ratio of majority class of minority class into 1:1, instead, to reduce the gap between the numbers of minority class and majority class. Different from CBOS, this method only cluster the majority class into $k$ clusters and regard each cluster as one subset of the majority class samples. After that combine each cluster with the whole minority class, and then the $k$ combined datasets will be considered as the updated training datasets. Finally, classify all the $k$ datasets with a classifier and choose the one that has the highest accuracy for building the training model.

**Majority Weighed Minority Over-sampling Technique (MWMOTE) (Barua et al. 2014):** This technique involves three key steps, identifies the most important and hard-to-learn minority class samples, $S_{imin}$, calculates a select weight $S_w$ from each member of $S_{imin}$, and generates the synthetic samples from $S_{imin}$ using $S_w$ and produce the output set $S_{omin}$ by adding the new generated samples to the original minority class, $S_{min}$.

Precisely, there are three stages in constructing $S_{imin}$. In the first stage, MWMOTE filters the original minority class samples, $S_{min}$, in order to find a filtered minority set, $S_{min f}$. In this respect, the nearest neighbour of each sample $x_i$ of $S_{min}$ is calculated, $NN(x_i)$. Then $x_i$ will be removed if its $NN(x_i)$ contains only the majority class samples. In the second stage, construct a borderline majorities, $N_{maj}(x_i)$ for each $x_i$ with expected the number of majority neighbours used for constructing informative minority samples ($k2$) as small as possible, and a borderline majority set $S_{bmaj}$ is obtained by combining all the $N_{maj}(x_i)$. Finally, MWMOTE constructs $N_{min}(y_i)$ regarding each $y_i \in S_{bmaj}$, and then we can obtain $S_{imin}$ by combining all the $N_{min}(y_i)$s.

For the weights of MWMOTE, $S_w$ is expressed as:

$S_w(x_i) = \sum_{y_i \in S_{bmaj}} I_w(y_i, x_i)$ where $I_w(y_i, x_i)$ is the information weight, which is computed as the product of the closeness factor, $C_f(y_i, x_i)$ and the density factor $D_f(y_i, x_i)$:

$$I_w(y_i, x_i) = C_w(y_i, x_i) \times D_w(y_i, x_i)$$

While the closeness factor $C_f(y_i, x_i)$ is defined as:

$$C_f(y_i, x_i) = \frac{f\left(\frac{1}{d_n(y_i, x_i)}\right)}{C_f(th)} \times CMAX$$

Where $C_f(th)$ and $CMAX$ are the user defined parameters and $f$ is a cut-off function which is:

$$f(x) = \begin{cases} x & \text{if } x \leq C_f(th) \\ C_f(th) & \text{otherwise} \end{cases}$$

Moreover, MWMOTE computes $D_f(y_i, x_i)$ by normalizing $C_f(y_i, x_i)$, which is:

$$D_f(y_i, x_i) = \frac{C_f(y_i, x_i)}{\sum_{q \in S_{imin}} C_f(y_i, x_i)}$$

In addition, in synthetic samples generating process, MWMOTE first cluster $S_{min}$ into $M$ clusters, which can be denoted as $L_1, L_2, \cdots, L_M$. Then select a sample $x$ from $S_{imin}$ following the probability distribution $\{S_p(x_i)\}$ (where $S_p(x_i) = S_w(x_i) / \sum_{z_i \in S_{imin}} S_w(z_i)$), let's assume $x \in L_k$. After that randomly choose another sample $y$ from $L_k$, and generate a synthetic sample $s$ using the linear interpolation of $x$ and $y$, which is:

$$s = x + \alpha \times (y - x)$$

Where $\alpha$ is a random number of $[0,1]$.

**Feature Selection (FS) (Yu et al. 2014):** This technique is employed in the asBagging_FSS method (Yu et al. 2014), in which the irrelevant and redundant features are expected to be removed. In order to delete the redundant features, FS first collect the similar features into multiple different groups by using hierarchical clustering method which utilizes Pearson correlation coefficient (Wang et al. 2005). The Pearson correlation coefficient computes the similarity across two features $f_i$ and $f_j$ as:

$$Sim(f_i, f_j) = \frac{\sum_{k=1}^{N}(f_{ik} - \overline{f_i})(f_{jk} - \overline{f_j})}{\sqrt{\sum_{k=1}^{N}(f_{ik} - \overline{f_i})^2}\sqrt{\sum_{k=1}^{N}(f_{jk} - \overline{f_j})^2}}$$

Where $f_{ik}$ is the value of $f_i$ on the $k$th sample, $\overline{f_i}$ is the mean of $f_i$ and $N$ is the size of the training dataset. After the multiple clusters are obtained, FS utilize the following signal-noise ratio (SNR) to extract the most relative features of the classification task:

**TABLE 1: Biomedical data used in this study.**

| Dataset | Abbre. | #Samples | #Genes | #Min | #Maj | IR | Distribution |
|---|---|---|---|---|---|---|---|
| Leukemia | Leu | 72 | 7129 | 25 | 47 | 1.34 | Skellam |
| Ovarian | Ova | 253 | 15154 | 91 | 162 | 1.78 | Generalized Extreme Value |
| Lung-cancer (Dana-Farber Cancer) | LCD | 203 | 12600 | 64 | 139 | 2.17 | T Location-Scale |
| Central Nervous System | CNS | 60 | 7129 | 21 | 39 | 1.86 | Skellam |
| DLBCLT | DLBCLT | 77 | 6817 | 19 | 58 | 3.05 | Skellam |
| Lung cancer (Brigham and - Women's Hospital) | LCB | 181 | 12533 | 3S1 | 150 | 4.84 | T Location-Scale |
| E2A-PBX1 | EP1 | 327 | 12599 | 27 | 300 | 11.11 | Generalized Pareto |
| BCR-ABL | BCL | 327 | 12599 | 15 | 312 | 20.80 | Generalized Pareto |
| Golub-1999-v1 | G99v1 | 72 | 1868 | 25 | 47 | 1.88 | Negative Binomial |
| Golub-1999-v2 | G99v2 | 72 | 1868 | 25 | 47 | 1.88 | Negative Binomial |
| Armstrong-2002-v1 | A02v1 | 72 | 1081 | 24 | 48 | 2.00 | Generalized Pareto |
| Armstrong-2002-v2 | A02v2 | 72 | 2194 | 24 | 48 | 2.00 | Generalized Pareto |
| Dyrskjot-2003 | D03 | 40 | 1203 | 11 | 29 | 2.64 | Generalized Extreme Value |
| Pomeroy-2002-v1 | P02v1 | 34 | 857 | 9 | 25 | 2.78 | Negative Binomial |
| Shipp-2002-v1 | S02v1 | 77 | 798 | 19 | 58 | 3.05 | Negative Binomial |
| Singh-2002 | S02 | 69 | 339 | 16 | 50 | 3.13 | Generalized Pareto |
| Pomeroy-2002-v2 | P02v2 | 42 | 1379 | 10 | 32 | 3.20 | Negative Binomial |
| Yeoh-2002-v1 | Y02v1 | 248 | 2526 | 43 | 205 | 4.77 | Generalized Pareto |
| Gordon-2002 | G02 | 181 | 1626 | 31 | 150 | 4.84 | Generalized Extreme Value |
| Su-2001 | S01 | 174 | 1571 | 26 | 148 | 5.69 | Negative Binomial |

$$SNR(f_i) = |\mu_0 - \mu_1|/(\sigma_0 + \sigma_1)$$

Where $\mu_0$ and $\mu_1$ are mean values of feature $f_i$ belonging to two different classes, $\sigma_0$ and $\sigma_1$ are their standard deviations. According to (Wang et al. 2005), the extracted features are closely related to classification and approximatively non-redundant. We employ this technique specifically, because it is a recently developed method, and performs well (Yu et al. 2014).

## 3 DATASET

Table 1 describes the information about publicly available DNA datasets used in this paper. The first column is the dataset name, the second column is the abbreviation form of the dataset, which will be used in later section. The last second column describes the imbalance ratio (IR), and the last column is the identified distribution of each dataset, which will be used for further analyzing the distribution-based discussion. The detailed information about the datasets distribution can be seen from Fig.1. Moreover, the first eight datasets are publicly available at http://datam.i2r.a-star.edu.sg/datasets/krbd/, and the other datasets are publically available at https://github.com/dcyoung/MultinomialRegression/tree/master/src/data.

**Fig. 1: Datasets Distribution Display.**

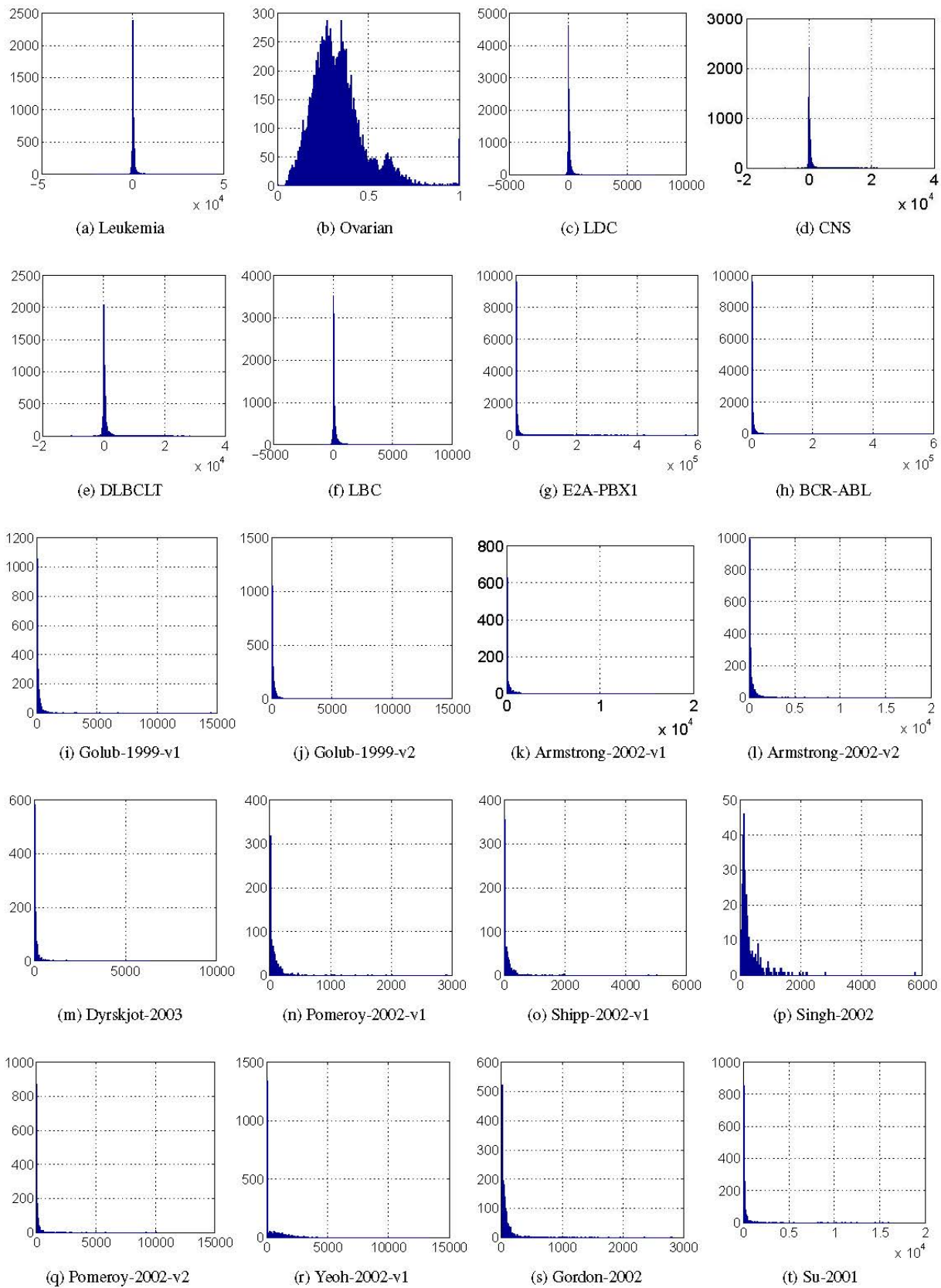

$$IR = \frac{\#\text{majority class samples}}{\#\text{minority class samples}} \qquad (2)$$

**Leukemia dataset**, also spelled leukaemia, developed in 352,000 people globally and caused 265,000 deaths in 2012. And the Leukemia gene expression profiling dataset is widely used as a way of disease diagnosis (Schuler et al. 2014). The dataset used in this paper contained 72 bone marrow samples including 47 acute lymphoblastic leukemia (ALL) and 35 acute myeloid leukemia (AML). Each sample contains 7129 genes.

**Ovarian cancer dataset (Petricoin et al. 2002):** this dataset consists of 91 controls (normal) and 162 ovarian cancers. The raw dataset of each sample contains the relative amplitude of the intensity at each molecular mass/ charge (M/Z) identity. The dataset used in this study is the one that normalized. The normalization is done over all the 253 samples for all the 15154 M/Z identities.

**Lung Cancer (Dana-Farber Cancer Institute, Harvard Medical School) (Bhattacharjee et al. 2001) dataset:** this dataset includes a total of 203 snap-frozen lung tumors and normal lung samples. The 203 speciments contain 139 samples of lung adenocarcinomas, 21 samples of squamous cell lung carcinomas, 20 samples of pulmonary carcinoids, 6 samples of small-cell carcinomas and 17 normal lung samples. For each sample, there are 12600 genes. In this study, we set the 139 samples of lung adenocarcinomas, as majority class and the other samples as the minority class.

**The central nervous system (CNS) dataset:** this dataset is the part of the nervous system consisting of the brain and spinal cord. The authors (Pomeroy et al. 2002) studied this disease by developing a classification system based on DNA microarray gene expression data, and demonstrated that the clinical outcome of children with medulloblastomas is highly predictable on the basis of the gene expression profiles of their tumours at diagnosis. Moreover, another study (Hajj-Ali and Calabrese 2014) further researched on the problem and pointed out that great work is still need to expatiate on different aspects of CNS vasculitis. The dataset used in this paper contains 60 samples of which 39 are patient samples and 21 are survivors. There are 7129 genes in the dataset.

**DLBCLT data Diffuse large B-cell lymphoma (Harvard Medical School & Whitehead MIT) dataset:** this dataset is the most common lymphoid malignancy in adults, and its curable is less than 50% (Shipp et al. 2002). The dataset chosen in this study contains 77 samples with 58 DLBCL and 19 Follicular Lymphoma (FL) samples. Each sample has 6817 genes.

**Lung cancer (Brigham and Women's Hospital, Harvard Medical School) dataset:** this dataset is the leading cause of cancer death in United States. According to Bhattacharjee et al., gene expression profiling can be used for lung cancer prediction (Lin et al. 2013). And researchers (Yu et al. 2013) have studied this problem from data mining perspective . The lung cancer data used in this paper contains 181 tissue sample of which 31 are malignant pleural mesothelioma (MPM) and 150 are adenocarcinoma (ADCA). Each sample is composed of 12533 genes.

**Leukemia (Stjude data) dataset (Yeoh et al. 2002):** this data has been divided into six diagnostic groups (BCR-ABL, E2A-PBX1), and one that contains diagnostic samples that did not fit into any one of the above groups (labelled as "Others"). There are 12558 genes. In our study, for datasets are chosen from Leukemia (Stjude data): BCR-ABL, E2A-PBX1, Hyperdip50 and TEL-AML1. For each dataset, we set the class that labelled with 'Others' as the majority class, while the other class as the minority class.

**Golub-1999-v1 dataset (Su et al. 2001):** Golub-1999 dataset was originally collected for the purpose of automatically discover the distinction between acute myeloid leukemia (AML) and acute lymphoblastic leukemia (ALL) without previous knowledge of these classes. Further, examination results showed that the procedure can further categorize distinguish between B-cell and T-cell ALL. Golub-1999-v1 includes 72 samples of which 47 samples were labelled as ALL, and 25 samples were labelled as AML, each sample has 1877 genes, while Golub-1999-v2 (short for G1999v2) composes 72 samples with 38 samples labelled ALL-B, 9 samples were labelled as ALL-T and 25 samples were labelled as AML. Each sample is described by 1877 genes. In this study we set the AML class as the minority class, while the other two class as the majority class.

**Armstrong-2002-v1 dataset (Su et al. 2001):** this dataset includes two classes, that is, 24 minority class samples (which were labelled as ALL) and 48 majority class samples (MLL). The dataset used in this study is described by 1081 genes. And Armstrong-2002-v2 s(short for A2002v2) has been divided into three groups (24 samples were labelled as ALL, 20 samples were labelled as MLL and 28 samples were labelled as AML). There are 2194 genes in each sample.

**Dyrskjot-2003 dataset (Su et al. 2001):** this dataset which is a common malignant disease, contains 40 samples and 1203 genes. The dataset is divided into three groups with each group contains 9 (T2+), 20(TA), 11(T1) samples. In this study, we set the T1 class as the minority class and the other two classes as the majority class.

**Pomeroy-2002-v1 dataset (Su et al. 2001):** this data is about medulloblastomas (MD) tumor, which is the most common malignant brain tumor of childhood. These DNA microarray gene expression data are derived from 99 patient samples which were demonstrated that medulloblastomas are molecularly distinct from other brain tumors including primitive neuroectodermal tumours (PNET), atypical teratoid/rhabdoid tumours (Rhab) and malignant gliomas (Mglio). Within the class of medulloblastomas (MD), they also studied the heterogeneity of classic (C) desmoplastic (D) ones. In the reported data, normal tissues were also considered (Ncer). Pomeroy-2002-v1 data include 25 CMD samples and 9 DMD samples, while Pomeroy-2002-v2 (short for P2002v2) include 10 MD, 10 Mglio, 10 Rhab, 4 Ncer and 8 PENT samples.

**Shipp-2002-v1 dataset (Su et al. 2001):** this data includes two malignancies, which are 58 B-cell lymphoma (DLBCL) samples and 19 related GC B-cell lymphoma, follicular (FL) samples. In this study, each sample is described by 798 genes.

**Singh-2002 dataset dataset (Su et al. 2001):** The original data include 52 prostate tumors (PR) and 50 Normal prostate specimens (N). For this study, we randomly selected 16 samples from 52 prostate tumors as the minority class.

**Yeoh-2002-v1 dataset dataset (Su et al. 2001):** this data is originally collected for determining whether gene expression profiling could enhance risk assignment. In Yeoh-2002-v1 dataset, there are 43 T-ALL and 205 B-ALL samples.

**Gordon-2002 dataset dataset (Su et al. 2001):** this dataset is composed of 31 malignant pleural mesothelioma (MPM) samples and 150 adenocarcinoma (AD) samples of the lung. And each sample is described by 1626 genes in this study.

**Su-2001 (Su et al. 2001):** This dataset describes the RNA profiling for carcinomas of the prostate (PR), breast (BR), lung (LU), ovary (OV), colorectum (CO), kidney (KI), liver (LI), pancreas (PA), bladder/ureter (BL), and gastroesophagus (GA), which collectively account for approximately 70% of all cancer-related deaths in the United States. In the experiment, we set the RP samples as the minority class, and all other samples belong to the majority class.

# 4 EXPERIMENTAL FRAMEWORK

In this section, we present the parameters setting for the experiment, the classification algorithms considered and evaluation criteria using in the study.

## 4.1 Parameters Setting

In the comparative study, it is very necessary to set the baseline study for the performance evaluation. Therefore, the original (ORI) datasets without applying any sampling methods, are used to provide a baseline for the performance evaluation.

In addition, the parameters in MWMOTE are set as $k1 = 5$, $k2 = 3$, $k3 = S_{min}/2$, $C_f(th) = 5$, and $CMAX = 2$. For WE, CBOS and CBUS, $k$ is set to three, five and three, respectively. Regarding feature selection (FS) technique, the dimension of feature space is initially designated as 100 which is also initially recommended by (Yu et al. 2014).

The undersampling rate for RUS is set to 50%, which indicates the percentage of the majority class to be removed. For example, if a dataset contains 1000 majority class instances, '50%' means after resampling, 50% that is 500 majority instances will be removed. According to (Barua et al. 2014), the oversampling rate for ROS and MWMOTE is set as 200%. '200%' mean the minority class size will be doubled after applying each oversampling approach. For all other techniques, including classification algorithms (with $k = 1$ for $k$NN classifier), we use the default parameters settled with Matlab2015 research version.

## 4.2 Classification Methods

Five classifiers: C4.5 decision tree, Support Vector Machine (SVM), K-Nearest Neighbors (KNN), Linear Discriminant Analysis (LDA) and Random Forests (RF) which are previously employed in imbalanced biomedical study (Zhao et al. 2008), (Khoshgoftaar et al. 2015), (Dittman et al. 2015), are considered in this work.

**C4.5 decision tree (Quinlan 2014),** which builds decision trees using an entropy-based splitting criterion stemming, is the very sensitive to class imbalances. This is because C4.5 works globally, not paying attention to specific data points. C4.5 as a learning algorithm, improves upon ID3 by adding support for handling missing values and tree pruning.

**SVM(Cortes and Vapnik 1995)** is a classifier that for binary classification, which attempt to find out a linear combination of the variables that best divides the samples into two groups by constructing a hyperplane or set of hyperplanes in a high-dimensional space. The idea of separation is that the optimal linear combination of the variables can maximize the distance between the classes. However, when the perfect separation is not possible, the optimal linear combination will be determined by a criterion in order to minimize the number of misclassifications.

$k$NN **(Fix and Hodges Jr 1951)** is one of the most popular non-parametric classification approaches that classifies a new specimen based on the class labels of its nearest neighbours. And the class of the new specimen is predicted as the majority class label of its $k$ nearest neighbours ($k$ is a positive integer). If $k = 1$ then the object is simply assigned to the class of its nearest neighbour.

**TABLE 2: Confusion matrix.**

|  | Predicted Positive | Predicted Negative |
|---|---|---|
| Actual Positive | True positive(TP) | False negative(FN) |
| Actual Negative | False positive(FP) | True negative(TN) |

**LDA (Wang and Tang 2004)** is the most commonly used as dimensionality reduction technique in the data processing step in the application of pattern classification or machine learning. Its goal is to project a dataset into a lower-dimensional in order to find a set of projecting vectors that best discriminating different classes, and avoid overfitting in order to reduce the computation costs.

**RF (Breiman 2001)** is a generalization of standard decision trees, based on bagging from a single training set of random not pruned decision tree. And a majority vote is utilized for the final decision make for a given observation. Precisely, during the training process, about one-third of the training data are not used in creating the decision tree model; these training data are called out-of-bag (OOB), which is used for testing the model and generating an unbiased estimator of the error rate. In this respect, there is no need to provide a set of additional tests or cross-validation to evaluate the model. Moreover, RF was used for class imbalance learning in predicting customer profitability and retention (Larivière and Van den Poel 2005).

We report the average of 50 runs of each experiment in which the datasets are randomly partitioned into the training data and the testing data. We use 60% of the whole data as training data and the remaining 40% for testing in our study. That is to say, we have created $20 \times 50 \times 8 \times 5 = 40000$ models in the experimental process (20 stand for the number of datasets, 50 means run times, 8 for eight data pre-processing techniques, 5 for number of classification algorithms ), and only averaged results and the standard deviation are reported.

### 4.3 Evaluation Criteria

Generally, the minority class is labelled as the positive class and the majority class is marked as the negative class. The confusion matrix values are true positive (TP), false positive (FP), true negative (TN), and false negative (FN) (Table 2). It is well-known that in the imbalanced data, the overall accuracy usually biased toward the majority class, thus some other specific evaluation metrics (Cano et al. 2013), such as Precision (Pre), Recall, F-measure (FM) and area under the receiver operating characteristic curve (AUC) (Galar et al. 2013) are used as supplementary evaluation criteria. F-measure which outputs a single value reflecting the 'goodness' of the classification performance with minority class, is defined as the harmonic mean of recall and precision. AUC which is not sensitive to the distribution between the majority and minority classes, can sort models by overall performance, and thus is more considered in models assessment. Based on Table 2, these performance measures are calculated as follows:

$$\text{Acc} = \frac{TP+TN}{TP+FN+FP+TN}$$

$$\text{Rec} = \frac{TP}{TP+FN}$$

$$\text{FM} = \frac{(1+\beta) \times \text{Rec} \times \text{Pre}}{\beta^2 \times (\text{Rec+Pre})}$$

Where $\text{Pre} = \frac{TP}{TP+FP}$.

**Fig. 2: FM values in terms of C4.5, SVM, 1NN, LDA and RF classifiers.**

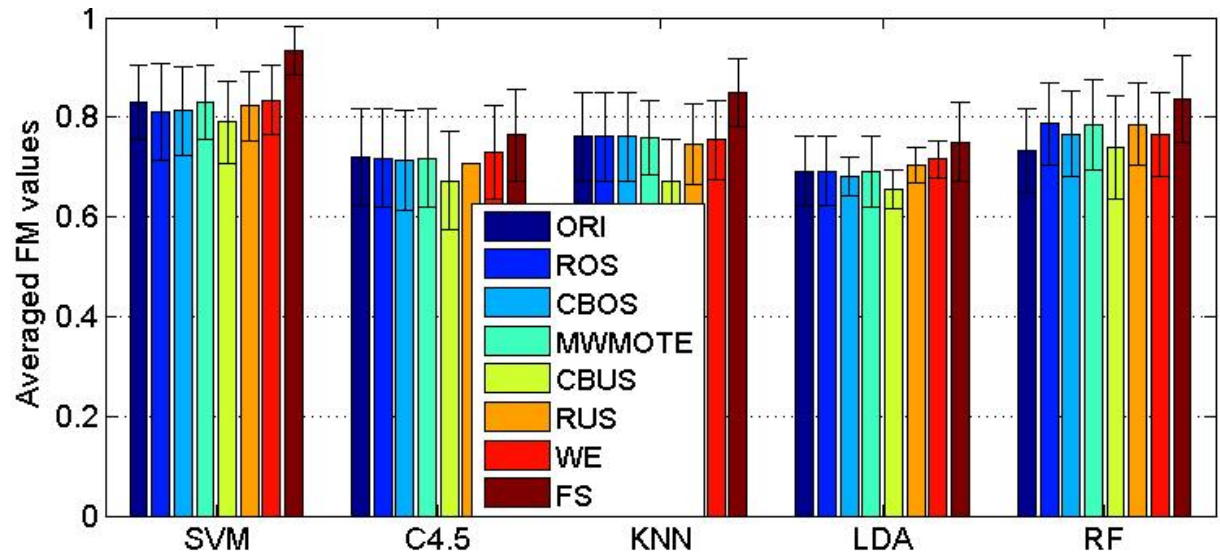

## 5 RESULTS AND DISCUSSION

Firstly, we present the average FM and AUC values in terms of the five classifiers. Then a detail discussion from data distribution perspective will be presented.

### 5.1 Overall View

In this section, we will provide an overall view of the performance of each technique in terms of the five classifiers. Fig.2 and Fig.3 describe the averaged FM and AUC values of the eight methods using five classifiers across all the twenty datasets.

Precisely, from Fig.2 one can see that among all the classifiers, techniques that using SVM classification algorithm performs the best, followed by $k$NN and RF, and Feature Selection (FS) exhibits the best averaged performance in all scenarios with at least 5% higher FM and AUC values (except for C4.5 classifier) than ORI. Comparatively, resampling techniques that with C4.5 and LDA classifiers result in worse performance. Fig.3 also indicates similar results.

Fig.2 and Fig.3 also present that not all resampling techniques can significantly improve the imbalanced biomedical data learning. For example, when using KNN classifier, we can see that most methods (except FS) have not shown better results, specifically, CBUS even performs far worse with nearly 8% less than ORI. Similarly, resampling techniques such as ROS, CBOS, CBUS and RUS do not exhibit any significant performance when using SVM learner. Most of resampling approaches have shown better performance with RF learner. For instance, ROS, MWMOTE and RUS obtain about 6% higher FM value than ORI, meanwhile CBOS and WE result in nearly 3.5% higher FM value compared to ORI. In addition, it is clear to observe that most resampling techniques with SVM classifier can achieve at least 4% higher FM and AUC values compared to C4.5, KNN, and LDA. Based on the above analysis, it is worthwhile to point out that we only report the results that are obtained by using SVM, KNN and RF classifiers in the following discussion because of their better performance in class imbalance biomedical data learning.

Fig. 3: AUC values in terms of C4.5, SVM, 1NN, LDA and RF classifiers.

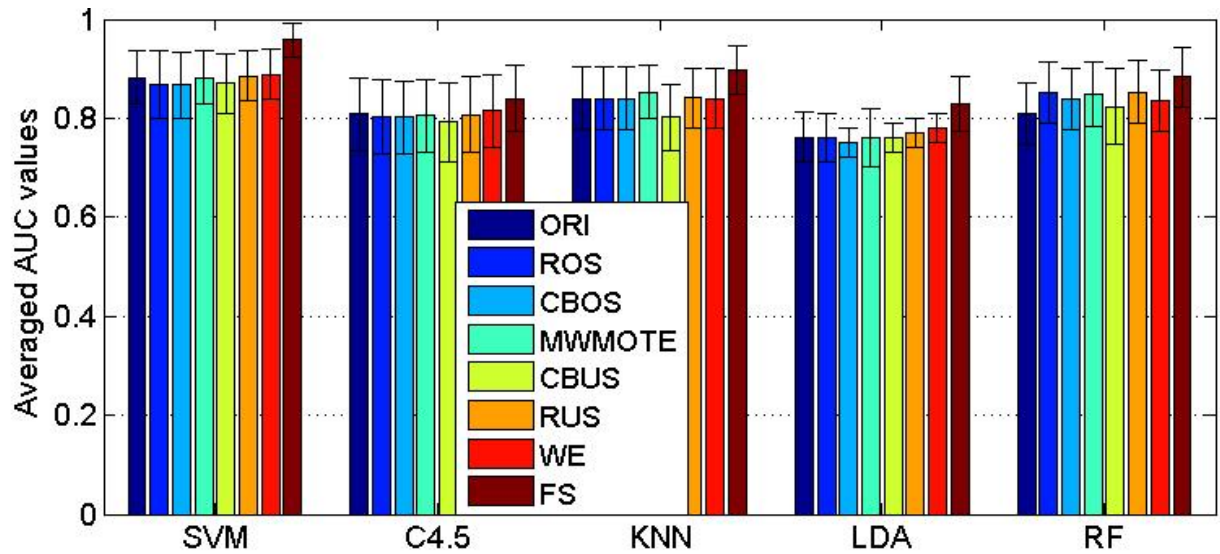

Fig. 4: Averaged performance in terms of SVM.

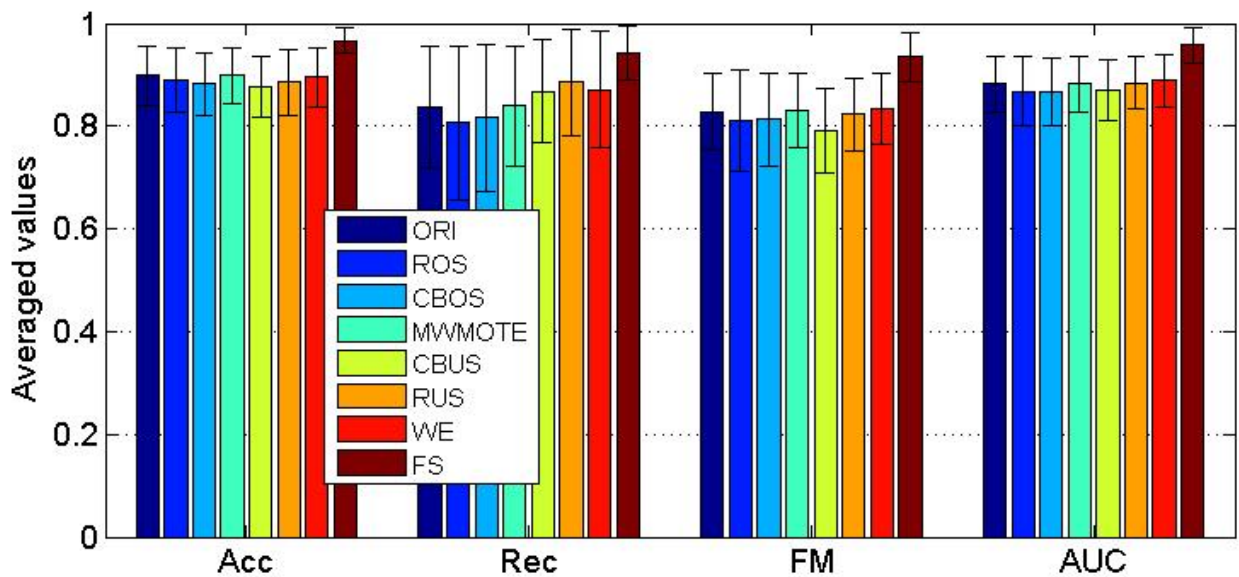

## 5.2 Classification Performance in terms of Data Distributions

### 5.2.1 Results of SVM Classifier

Fig.4 reports the averaged performance of eight resampling techniques in terms of SVM classifier, four performance measures on 20 datasets. It can be observed that FS outperforms all the other techniques with at least 7% higher for Acc, Rec, FM and AUC, while other techniques have not shown stable results across all the performance measures. For example, CBUS and RUS appear better performance with Rec, however, the performance of FM and Acc is even worse than ORI. Interestingly, although WE has not shown outstanding performance, it appears to be quite stable compared to the other resampling methods with better Rec, FM and AUC performance than ORI.

Table 3 presents the detailed results of the eight techniques on 20 datasets from the distribution-based perspective in terms of SVM classifier. One can see that FS results in higher FM and AUC values on Skellam distribution, Generalized Extreme Value distribution and Negative Binomial distribution datasets. For T Location-Scale and Generalized Pareto distribution datasets, FS also outperforms other methods in most cases, even though RUS achieves higher AUC values in two scenarios and WE obtains better results on Chowdary-2006 dataset, FS has achieved highly comparative results in such datasets. For example, RUS obtains FM values of 0.949 on Lung-cancer (Dana-Farber Cancer Institute) dataset, however, the FM value with FS technique is 0.945, which is only 0.004 lower.

**TABLE 3: FM and AUC performance with each technique across all the datasets in terms of SVM.**

| Distribution | dataset | ORI | | ROS | | CBOS | | MWMOTE | | CBUS | | RUS | | WE | | FS | |
|---|---|---|---|---|---|---|---|---|---|---|---|---|---|---|---|---|---|
| | | FM | AUC | FM | AUC | FM | AUC | FM | AUC | FM | AUC | FM | AUC | FM | AUC | FM | AUC |
| Skellam | Leu | 0.950 | 0.957 | 0.887 | 0.908 | 0.908 | 0.925 | 0.947 | 0.955 | 0.918 | 0.940 | 0.927 | 0.947 | 0.950 | 0.957 | **0.975** | **0.981** |
| | CNS | 0.421 | 0.574 | 0.413 | 0.569 | 0.413 | 0.575 | 0.441 | 0.580 | 0.424 | 0.537 | 0.502 | 0.579 | 0.490 | 0.582 | **0.896** | **0.925** |
| | DLBCLT | 0.906 | 0.932 | 0.902 | 0.928 | 0.909 | 0.934 | 0.910 | 0.936 | 0.812 | 0.907 | 0.873 | 0.934 | 0.893 | 0.944 | **0.966** | **0.984** |
| Generalized Extreme Value | Ova | 0.997 | 0.998 | 0.983 | 0.985 | 0.981 | 0.983 | 0.996 | 0.997 | 0.944 | 0.965 | 0.991 | 0.994 | 0.995 | 0.996 | **0.999** | **0.999** |
| | Dy03 | 0.528 | 0.661 | 0.508 | 0.649 | 0.480 | 0.625 | 0.535 | 0.667 | 0.585 | 0.730 | 0.557 | 0.699 | 0.573 | 0.694 | **0.926** | **0.947** |
| | G02 | 0.968 | 0.973 | 0.969 | 0.972 | 0.971 | 0.974 | 0.970 | 0.974 | 0.961 | 0.977 | 0.963 | 0.975 | 0.968 | 0.973 | **0.990** | **0.996** |
| T Location-Scale | LDC | 0.916 | 0.933 | 0.847 | 0.884 | 0.837 | 0.876 | 0.914 | 0.931 | 0.919 | 0.948 | **0.926** | **0.949** | 0.922 | 0.939 | **0.926** | 0.945 |
| | LBC | 0.972 | 0.973 | 0.972 | 0.973 | 0.966 | 0.968 | 0.974 | 0.975 | 0.982 | 0.983 | 0.984 | 0.985 | 0.972 | 0.973 | **0.992** | **0.998** |
| Generalized Pareto | H50 | 0.908 | 0.929 | 0.868 | 0.895 | 0.858 | 0.886 | 0.909 | 0.928 | 0.850 | 0.917 | **0.919** | **0.946** | 0.910 | 0.934 | 0.908 | 0.940 |
| | EP1 | 0.967 | 0.969 | 0.963 | 0.966 | 0.955 | 0.960 | 0.967 | 0.969 | 0.901 | 0.966 | 0.969 | 0.975 | 0.970 | 0.972 | **0.995** | **1.000** |
| | BCL | 0.151 | 0.526 | 0.139 | 0.522 | 0.112 | 0.510 | 0.143 | 0.523 | 0.181 | 0.619 | 0.213 | 0.553 | 0.149 | 0.526 | **0.848** | **0.904** |
| | C06 | 0.931 | 0.942 | 0.910 | 0.930 | 0.924 | 0.936 | 0.931 | 0.942 | 0.918 | 0.931 | 0.931 | 0.943 | **0.951** | **0.960** | 0.937 | 0.948 |
| | A02v1 | 0.951 | 0.956 | 0.933 | 0.940 | 0.944 | 0.950 | 0.950 | 0.955 | 0.864 | 0.906 | 0.919 | 0.945 | 0.955 | 0.960 | **0.990** | **0.993** |
| | A02v2 | 0.937 | 0.945 | 0.915 | 0.930 | 0.917 | 0.930 | 0.932 | 0.941 | 0.853 | 0.898 | 0.929 | 0.949 | 0.936 | 0.944 | **0.998** | **0.998** |
| | S02 | 0.831 | 0.900 | 0.824 | 0.892 | 0.797 | 0.876 | 0.820 | 0.891 | 0.729 | 0.861 | 0.788 | 0.886 | 0.749 | 0.860 | **0.831** | **0.906** |
| | Y02v1 | 0.926 | 0.934 | 0.892 | 0.906 | 0.848 | 0.874 | 0.907 | 0.916 | 0.908 | 0.927 | 0.919 | 0.939 | 0.926 | 0.934 | **0.933** | **0.950** |
| Negative Binomial | G99v1 | 0.907 | 0.925 | 0.882 | 0.906 | 0.892 | 0.912 | 0.917 | 0.933 | 0.881 | 0.917 | 0.914 | 0.939 | 0.919 | 0.940 | **0.977** | **0.986** |
| | G99v2 | 0.915 | 0.930 | 0.841 | 0.885 | 0.909 | 0.925 | 0.919 | 0.934 | 0.879 | 0.915 | 0.912 | 0.938 | 0.932 | 0.948 | **0.981** | **0.988** |
| | P02v1 | 0.659 | 0.781 | 0.704 | 0.809 | 0.726 | 0.819 | 0.691 | 0.799 | 0.557 | 0.716 | 0.605 | 0.765 | 0.657 | 0.783 | **0.907** | **0.943** |
| | S02v1 | 0.810 | 0.875 | 0.801 | 0.864 | 0.809 | 0.869 | 0.814 | 0.876 | 0.792 | 0.872 | 0.779 | 0.879 | 0.804 | 0.880 | **0.920** | **0.946** |
| | P02v2 | 0.791 | 0.857 | 0.806 | 0.864 | 0.821 | 0.873 | 0.809 | 0.867 | 0.588 | 0.739 | 0.662 | 0.786 | 0.787 | 0.869 | **0.916** | **0.961** |
| | S01 | 0.992 | 0.999 | 0.991 | 0.998 | 0.996 | 0.999 | 0.992 | 0.999 | 0.971 | 0.995 | 0.988 | 0.998 | 0.992 | 0.999 | **1.000** | **1.000** |

5.2.2 Results of KNN Classifier

Fig.5 depicts the averaged performance of the eight techniques in terms of KNN classifier ($k=1$). It can be seen that FS outperforms ORI with at least 4% higher for all performance measures. However, this is not true for other techniques. For example, MWMOTE, CBUS, RUS and WE result in much better Rec value with at least 6% higher than ORI, while its Acc value is worse than ORI, which means this method is not practical in real world problems. We believe this is most possibly because the minority class region is wrongly enlarged by erroneously generated synthetic minority class samples, which will lead the minority class region falling inside the majority class region.

Table 4 details the performance of eight techniques from distribution perspective in terms of KNN classifier. Different from SVM classifier, Table 4 demonstrates that both MWMOTE and FS are beneficial to Generalized Extreme Value distribution datasets. RUS and FS benefit T Location-Scale distribution datasets. And FS performs better in most of times for Skellam distribution, Generalized Pareto distribution and Negative Binomial Distribution datasets. Although MWMOTE and WE result in higher AUC values on Yeoh-2002-v1 and Chowdary-2006 dataset, FS can achieve very comparative results in such scenarios. Take Yeoh-2002-v1 dataset for example, MWMOTE obtains the highest FM (0.952) and AUC (0.959) values, however, FS results in 0.940 and 0.955, which is only 0.012 and 0,004 lower.

**TABLE 4: FM and AUC performance with each technique across all the datasets in terms of KNN.**

| Distribution | dataset | ORI | | ROS | | CBOS | | MWMOTE | | CBUS | | RUS | | WE | | FS | |
|---|---|---|---|---|---|---|---|---|---|---|---|---|---|---|---|---|---|
| | | FM | AUC | FM | AUC | FM | AUC | FM | AUC | FM | AUC | FM | AUC | FM | AUC | FM | AUC |
| Skellam | Leu | 0.853 | 0.884 | 0.853 | 0.884 | 0.853 | 0.884 | **0.878** | **0.915** | 0.824 | 0.867 | 0.887 | 0.914 | 0.852 | 0.883 | 0.862 | 0.883 |
| | CNS | 0.449 | 0.556 | 0.449 | 0.556 | 0.449 | 0.556 | 0.471 | 0.532 | 0.476 | 0.547 | 0.467 | 0.537 | 0.473 | 0.548 | **0.616** | **0.704** |
| | DLBCLT | 0.747 | 0.870 | 0.747 | 0.870 | 0.747 | 0.870 | 0.699 | 0.853 | 0.699 | 0.849 | 0.680 | 0.843 | 0.723 | 0.875 | **0.923** | **0.968** |
| Generalized Extreme Value | Ova | 0.894 | 0.913 | 0.894 | 0.913 | 0.894 | 0.913 | 0.901 | 0.928 | 0.749 | 0.807 | 0.875 | 0.906 | 0.891 | 0.911 | **0.986** | **0.987** |
| | Dy03 | 0.597 | 0.722 | 0.597 | 0.722 | 0.597 | 0.722 | **0.704** | **0.797** | 0.514 | 0.668 | 0.633 | 0.751 | 0.576 | 0.713 | 0.652 | 0.746 |
| | G02 | 0.918 | 0.956 | 0.918 | 0.956 | 0.918 | 0.956 | 0.920 | 0.965 | 0.869 | 0.948 | 0.891 | 0.954 | 0.905 | 0.955 | **0.972** | **0.994** |
| T Location-Scale | LDC | 0.857 | 0.898 | 0.857 | 0.898 | 0.857 | 0.898 | 0.772 | 0.857 | 0.782 | 0.863 | **0.862** | **0.909** | 0.861 | 0.901 | 0.865 | 0.902 |
| | LBC | 0.957 | 0.977 | 0.957 | 0.977 | 0.957 | 0.977 | 0.946 | 0.986 | 0.938 | 0.977 | 0.959 | 0.982 | 0.957 | 0.977 | **0.990** | **0.997** |
| Generalized Pareto | H50 | 0.723 | 0.815 | 0.723 | 0.815 | 0.723 | 0.815 | 0.728 | **0.874** | 0.477 | 0.715 | 0.715 | 0.842 | 0.755 | 0.848 | **0.802** | 0.866 |
| | EP1 | 0.902 | 0.958 | 0.902 | 0.958 | 0.902 | 0.958 | 0.846 | 0.976 | 0.322 | 0.819 | 0.798 | 0.961 | 0.856 | 0.953 | **0.990** | **0.999** |
| | BCL | 0.450 | 0.691 | 0.450 | 0.691 | 0.450 | 0.691 | 0.374 | 0.756 | 0.145 | 0.649 | 0.398 | 0.733 | 0.456 | 0.716 | **0.650** | **0.815** |
| | C06 | 0.940 | 0.950 | 0.948 | 0.955 | 0.948 | 0.955 | 0.964 | 0.969 | 0.944 | 0.953 | 0.950 | 0.959 | **0.963** | **0.969** | 0.941 | 0.955 |
| | A02v1 | 0.873 | 0.899 | 0.873 | 0.899 | 0.873 | 0.899 | 0.909 | 0.935 | 0.851 | 0.886 | 0.913 | 0.936 | 0.874 | 0.901 | **0.956** | **0.967** |
| | A02v2 | 0.880 | 0.902 | 0.880 | 0.902 | 0.880 | 0.902 | 0.894 | 0.926 | 0.815 | 0.862 | 0.874 | 0.904 | 0.880 | 0.902 | **0.986** | **0.990** |
| | S02 | 0.509 | 0.684 | 0.509 | 0.684 | 0.509 | 0.684 | 0.550 | 0.713 | **0.657** | **0.734** | 0.510 | 0.688 | 0.543 | 0.709 | 0.639 | 0.698 |
| | Y02v1 | 0.886 | 0.900 | 0.886 | 0.900 | 0.886 | 0.900 | **0.952** | **0.959** | 0.886 | 0.900 | 0.903 | 0.914 | 0.886 | 0.900 | 0.940 | 0.955 |
| Negative Binomial | G99v1 | 0.857 | 0.895 | 0.857 | 0.895 | 0.857 | 0.895 | 0.851 | 0.896 | 0.805 | 0.856 | 0.833 | 0.882 | 0.862 | 0.905 | **0.902** | **0.927** |
| | G99v2 | 0.876 | 0.912 | 0.876 | 0.912 | 0.876 | 0.912 | 0.865 | 0.909 | 0.818 | 0.868 | 0.835 | 0.885 | 0.874 | 0.913 | **0.913** | **0.937** |
| | P02v1 | 0.460 | 0.639 | 0.460 | 0.639 | 0.460 | 0.639 | 0.470 | 0.653 | 0.417 | 0.599 | 0.442 | 0.630 | 0.408 | 0.585 | **0.767** | **0.884** |
| | S02v1 | 0.455 | 0.651 | 0.455 | 0.651 | 0.455 | 0.651 | 0.508 | 0.685 | 0.456 | 0.652 | 0.513 | 0.691 | 0.479 | 0.667 | **0.576** | **0.726** |
| | P02v2 | 0.774 | 0.866 | 0.774 | 0.866 | 0.774 | 0.866 | 0.742 | 0.851 | 0.606 | 0.755 | 0.668 | 0.806 | 0.682 | 0.836 | **0.875** | **0.941** |
| | S01 | 0.999 | 0.999 | 0.999 | 0.999 | 0.999 | 0.999 | 1.000 | 1.000 | 0.999 | 0.999 | 0.999 | 0.999 | 0.999 | 0.999 | **1.000** | **1.000** |

**Fig. 5: Averaged performance in terms of KNN.**

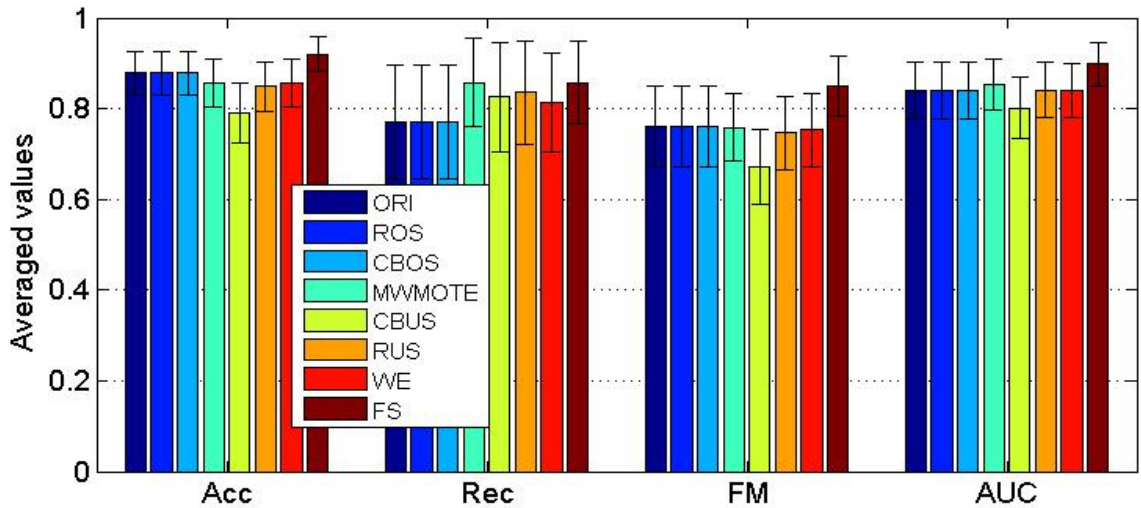

**Fig. 6: Averaged performance in terms of RF.**

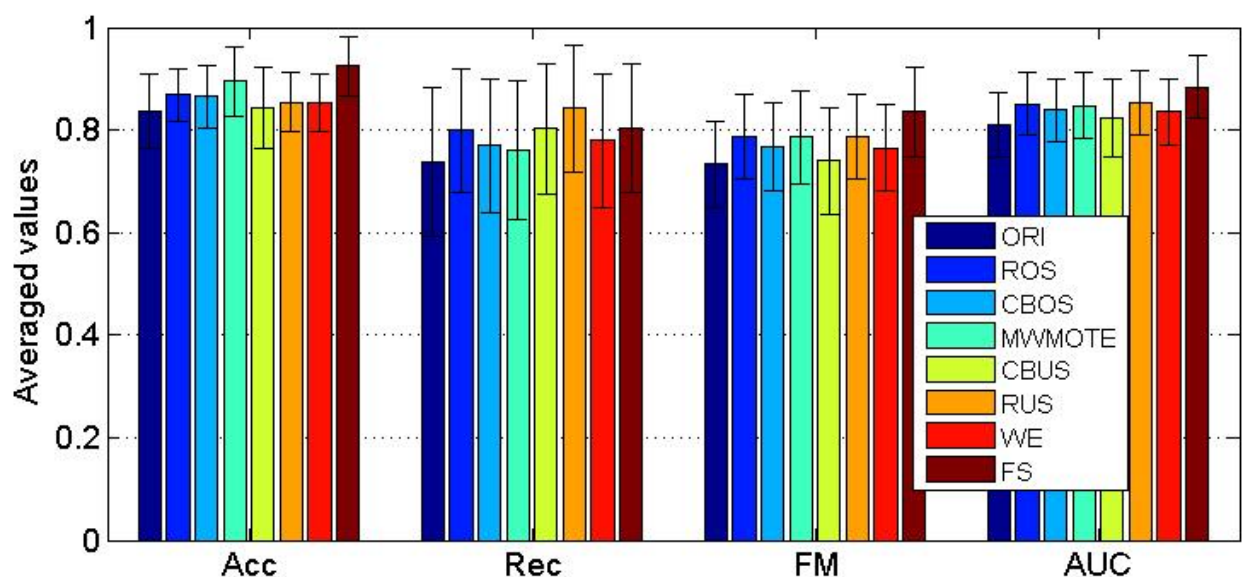

### 5.2.3 Results of RF Classifier

Fig.6 demonstrates the averaged performance of the eight techniques in terms of RF classifier. Interestingly, most resampling techniques exhibit significant improvement when using RF classifier. For example, ROS and RUS can achieve at least 2% higher Acc values and 4% other performance values than ORI. CBUS which appears performance the worst among all the resampling techniques, is also obtained 3% higher Rec value than ORI, and about 1% higher than ORI for Acc, FM and AUC values.

Table 5 illustrates the detailed performance of these techniques from data distribution perspective in terms of RF classifier. Obviously, we can see that when the imbalanced ratio is not severe, both WMWOTE and FS are beneficial to Generalized Pareto distribution and Negative Binomial distribution datasets. For all other scenarios, FS results in the highest FM and AUC values in all most all cases. For example, when facing Skellam distribution datasets, FS can obtain 0.700 AUC values, however, the second best performance comes from MWMOTE with 0.566, which is 13.6% lower than FS.

### 5.3 Discussion

From the aforementioned discussion, we can see that RUS and WE perform far worse in most of the times compared with FS. We believe the reason is being that RUS removes some very important information, especially when facing the small datasets, which will lead to insufficient information for training the classifier. Moreover, it has been shown that ROS has not led to significant improvement in all most all cases because of the potential overfitting (He et al. 2009). To be more specific, overfitting in oversampling occurs when classifiers produce multiple clauses in a rule for multiple copies of the same example which causes the rule to become too specific; even though ROS may result in higher accuracy values in training process, the classification performance on the unseen testing data is generally very poor. In this respect, CBOS which is also an oversampling method, faces a similar problem as ROS.

For other techniques, such as MWMOTE, the reason why these techniques are not stable for most of the time, is that the minority class region of the biomedical datasets is wrongly enlarged by erroneously generated synthetic minority class samples, which will lead the minority class region falling inside the majority class region. In addition, we can see that FS technique has resulted in a much better classification performance in most cases. We believe this is because FS has initially selected some highly essential and useful features, which means some redundant and irrelevant features have been cleaned up before training the classification model, thus the selected features could help the classifiers work more effectively.

**TABLE 5: FM and AUC performance with each technique across all the datasets in terms of RF.**

| Distribution | dataset | ORI | | ROS | | CBOS | | MWMOTE | | CBUS | | RUS | | WE | | FS | |
|---|---|---|---|---|---|---|---|---|---|---|---|---|---|---|---|---|---|
| | | FM | AUC | FM | AUC | FM | AUC | FM | AUC | FM | AUC | FM | AUC | FM | AUC | FM | AUC |
| Skellam distribution | Leu | 0.883 | 0.900 | 0.932 | 0.943 | 0.931 | 0.941 | 0.924 | 0.937 | 0.889 | 0.918 | 0.904 | 0.927 | 0.891 | 0.907 | **0.975** | **0.976** |
| | CNS | 0.297 | 0.517 | 0.429 | 0.571 | 0.365 | 0.539 | 0.430 | 0.566 | 0.400 | 0.506 | 0.470 | 0.532 | 0.383 | 0.515 | **0.574** | **0.700** |
| | DLBCLT | 0.586 | 0.721 | 0.722 | 0.811 | 0.715 | 0.805 | 0.740 | 0.821 | 0.652 | 0.786 | 0.726 | 0.827 | 0.675 | 0.795 | **0.807** | **0.857** |
| Generalized Extreme Value | Ova | 0.973 | 0.975 | 0.976 | 0.978 | 0.980 | 0.982 | 0.977 | 0.979 | 0.730 | 0.788 | 0.974 | 0.980 | 0.974 | 0.975 | **0.989** | **0.989** |
| | Dy03 | 0.419 | 0.567 | 0.583 | 0.683 | 0.495 | 0.649 | 0.450 | 0.596 | **0.580** | **0.727** | 0.540 | 0.689 | 0.455 | 0.595 | **0.597** | **0.701** |
| | G02 | 0.971 | 0.975 | 0.971 | 0.979 | 0.978 | 0.983 | 0.974 | 0.979 | 0.965 | 0.980 | 0.978 | 0.983 | 0.973 | 0.977 | **0.974** | **0.986** |
| T Location -Scale | LDC | 0.856 | 0.882 | 0.883 | 0.906 | 0.878 | 0.900 | 0.901 | 0.924 | 0.774 | 0.851 | 0.871 | 0.911 | 0.864 | 0.889 | **0.906** | **0.925** |
| | LBC | 0.963 | 0.966 | 0.979 | 0.982 | 0.978 | 0.980 | 0.968 | 0.971 | 0.978 | 0.985 | 0.974 | 0.979 | 0.970 | 0.973 | **0.984** | **0.988** |
| Generalized Pareto | H50 | 0.788 | 0.835 | 0.862 | 0.895 | 0.862 | 0.893 | **0.882** | **0.909** | 0.778 | 0.863 | **0.880** | **0.912** | 0.816 | 0.855 | **0.882** | **0.909** |
| | EP1 | 0.930 | 0.938 | 0.970 | 0.972 | 0.976 | 0.978 | **0.985** | **0.987** | 0.948 | 0.968 | 0.975 | 0.977 | 0.951 | 0.955 | **0.985** | **0.987** |
| | BCL | 0.088 | 0.500 | 0.088 | 0.500 | 0.096 | 0.503 | **0.387** | **0.628** | 0.187 | 0.549 | 0.092 | 0.502 | 0.088 | 0.500 | **0.387** | **0.628** |
| | C06 | 0.958 | 0.964 | 0.962 | 0.967 | 0.963 | 0.968 | **0.965** | **0.970** | 0.940 | 0.950 | 0.961 | 0.967 | 0.962 | 0.966 | **0.968** | **0.972** |
| | A02v1 | 0.961 | 0.966 | 0.971 | 0.974 | 0.969 | 0.972 | 0.971 | 0.973 | 0.949 | 0.962 | 0.956 | 0.968 | 0.959 | 0.963 | **0.999** | **0.999** |
| | A02v2 | 0.943 | 0.951 | 0.960 | 0.966 | 0.962 | 0.966 | 0.958 | 0.964 | 0.896 | 0.928 | 0.934 | 0.953 | 0.933 | 0.943 | **0.999** | **0.999** |
| | S02 | 0.667 | 0.785 | 0.696 | 0.809 | 0.641 | 0.770 | **0.710** | **0.825** | 0.662 | 0.813 | 0.686 | 0.808 | 0.670 | 0.795 | 0.674 | 0.794 |
| | Y02v1 | 0.730 | 0.799 | 0.738 | 0.800 | 0.534 | 0.688 | 0.594 | 0.720 | 0.725 | 0.797 | 0.789 | 0.842 | 0.701 | 0.781 | **0.922** | **0.938** |
| Negative Binomial | G99v1 | 0.905 | 0.921 | **0.943** | **0.954** | 0.920 | 0.933 | 0.934 | 0.946 | 0.905 | 0.934 | 0.921 | 0.942 | 0.920 | 0.934 | 0.901 | 0.918 |
| | G99v2 | 0.893 | 0.912 | **0.933** | **0.947** | 0.920 | 0.933 | 0.925 | 0.940 | 0.894 | 0.926 | 0.926 | 0.945 | 0.922 | 0.936 | 0.923 | 0.933 |
| | P02v1 | 0.504 | 0.622 | 0.571 | 0.677 | 0.568 | 0.690 | 0.556 | 0.687 | 0.514 | 0.659 | 0.545 | 0.687 | 0.524 | 0.673 | **0.722** | **0.801** |
| | S02v1 | 0.570 | 0.698 | 0.678 | 0.784 | 0.661 | 0.774 | 0.619 | 0.743 | 0.578 | 0.717 | **0.705** | **0.820** | 0.620 | 0.746 | **0.722** | **0.804** |
| | P02v2 | 0.438 | 0.584 | 0.600 | 0.724 | 0.582 | 0.698 | 0.550 | 0.688 | 0.562 | 0.706 | 0.677 | 0.769 | 0.579 | 0.696 | **0.687** | **0.775** |
| | S01 | 0.980 | 0.981 | 0.983 | 0.984 | 0.983 | 0.984 | 0.982 | 0.983 | 0.980 | 0.981 | 0.978 | 0.979 | 0.978 | 0.979 | **0.985** | **0.986** |

With the breadth of secondary data becoming more available, we believe machine learning techniques will become more important in evaluating the internal consistency, reporting, replication, and reproducibility of studies. Actually, many researchers and communities have emerged using machine learning-based techniques for disease diagnosis and cancer prediction (Golub et al. 1999), (Nutt et al. 2003), (Khoshgoftaar et al. 2014), (Conrads et al. 2003), (Lian et al. 2016), (Krawczyk et al. 2016). For example, Golub et al. (Golub et al. 1999) realized there was no general approach for identifying new cancer classes, and developed a generic approach for cancer classification using DNA microarrays. In (Nutt et al. 2003), gene expression profiling coupled with class prediction approaches were investigated and identified to be more objective, explicit and consistent than standard pathology in classifying high-grade gliomas. Considering the advent of proteomics could be helpful in discovering novel biomarkers in terms of diagnosing diseases. Conrads et al. proposed a revolutionary approach in proteomic pattern recognition for early diagnosis of diseases such as ovarian cancer. Further, feature selection and randomly undersampling techniques have been studied in (Khoshgoftaar et al. 2014) using seven biomedical datasets . The results indicate that optimal approach depends on the choice of class ratio. Recently, Lian et al. (Lian et al. 2016) proposed a prediction system for PET imaging based cancer treatment outcome prediction using radiomic features extracted from FDG-PET images. The presented system aims to improve the prediction accuracy, and reduce the imprecision and overlaps between the binary classes. Experiments have emphasized the effectiveness of the prediction systems. In (Krawczyk et al. 2016), a complete, fully automatic and efficient clinical decision support system which using both image processing and EUSBoost classifier, has been proposed for breast cancer malignancy grading. However, biomedical data in general have class imbalance problem (Lin et al. 2013).

**TABLE 6: Techniques in terms of different classifiers.**

| Classifier | Technique |
|---|---|
| SVM | WE, FS |
| kNN | MWMOTE, FS |
| RF | ROS, MWMOTE, RUS, FS |

**TABLE 7: Techniques in terms of different distributions.**

| Distribution | Classifier | | |
|---|---|---|---|
| | SVM | 1NN | RF |
| Skellam | FS | FS | FS |
| Generalized Extreme Value | FS | MWMOTE, FS | FS |
| T Location -Scale | RUS, FS | RUS, FS | FS |
| Generalized Pareto | FS | FS | MWMOTE, FS |
| Negative Binomial | FS | FS | ROS (IR<2), FS |

### 5.4 Recommendations

Table 6 summarizes the relationship among different classifiers and techniques based on the foregoing study, while Table 7 outlines the benefits of the data pre-processing techniques from data distribution perspective. From aforementioned results, Table 6 and Table 7, we can conclude:

For sampling techniques, when using SVM classifier, WE and FS are better choices, while MWMOTE and FS are better choices in terms of KNN classifier. For RF classifier, ROS, MWMOTE RUS and FS can be considered when facing imbalanced learning problem.

Considering the different distribution of the data in terms of SVM classifier, FS is a good choice when facing all the five kinds of distributions studies in this paper, and when facing the specifically T Location-Scale distribution, RUS is also considerable.

Considering different distributions of the data in terms of KNN classifier, FS could be the first choice, while MWMOTE and RUS are also considerable when facing Generalized Extreme Value distribution and T Location-Scale distribution datasets.

Furthermore, when using RF classifiers, we recommend FS techniques as the first choice for all the five kinds of distributions, and MWMOTE is also another better choice in terms of special Generalized Pareto distribution.

However, one can see that ROS can only be considered with Negative Binomial distribution data when the imbalance ratio is not severe (no more than 2).

Lastly, when choosing classification algorithms for imbalanced biomedical data learning problem, we recommend SVM classifier as the first choice, and then KNN and RF classifiers are also considerable, while it is necessary to avoid C4.5 classifier.

# 6 CONCLUSION

In this paper, we have reviewed and evaluated some newly developed and most important methodologies for imbalanced biomedical data learning problem. Extensive experiment study has been conducted using five classifiers, eight imbalanced data learning techniques on 20 real-world datasets with four performance measures. Meanwhile, experimental results have been discussed based on different data distributions. To our knowledge, no previous work have been analyzed the relationship between data pre-processing techniques (such as resampling and feature selection methods) and the distributions of datasets. Experimental results demonstrate that resampling techniques cannot dramatically improve the imbalanced biomedical data learning problem when using SVM, KNN, RF and C4.5 classifiers. However, when using RF classifiers, most resampling techniques can improve the classification performance significantly. Experimental results also exhibit that most techniques appear better performances when using SVM classifier. Most importantly, we find that FS could be the best choice when conducting imbalanced biomedical data learning. Further, experimental results from distribution-based analysis reveal that FS benefits all the five kinds of distribution datasets in terms of SVM, KNN and RF classifiers, while RUS is beneficial to T Location-Scale distribution datasets in terms of SVM and KNN learners. MWMOTE is considerable for Generalized Extreme Value distribution and Generalized Pareto distribution datasets in terms of KNN and RF classification algorithms, respectively.